\pdfoutput=1
\documentclass[11pt,letterpaper]{article}

\usepackage[utf8]{inputenc}
\usepackage[T1]{fontenc}
\usepackage{amsmath,amssymb,amsfonts}
\usepackage{graphicx}
\usepackage{booktabs}
\usepackage{hyperref}
\usepackage{url}
\usepackage{xurl}
\usepackage[numbers]{natbib}
\usepackage[expansion=false]{microtype}
\usepackage{geometry}
\usepackage{float}
\usepackage{enumitem}
\usepackage{xcolor}

\graphicspath{{figures/}}

\title{
  Modeling Local Exploit Hazard\\
  \large A Bayesian Framework for Quantifying Exploit Risk and
  Operational Efficiency
}

\author{
  Stephen Shaffer\\
  \normalsize stephen@shaffer.systems
  \and
  Laura Voicu\\
  \normalsize laura.voicu@me.com
}

\begin{document}

\maketitle

% ===========================================================================
\begin{abstract}
This paper presents a \emph{local exploit hazard model}: a Bayesian framework
that converts the global probabilities produced by an exploit likelihood model
(ELM), such as the Exploit Prediction Scoring System (EPSS), into a daily
exploit hazard rate for an organization's own assets.
The model measures the exploit-prevention effectiveness of deployed controls
as a probability distribution. That distribution is seeded from a
subject-matter-expert opinion pool and updated through Beta-Binomial inference
from telemetry, breach-and-attack simulation, or penetration testing, then
applied to ELM scores by attack-vector alignment.
The resulting per-vulnerability exploitation likelihoods are converted into
hazard rates using standard survival-analysis techniques, supporting both a
constant exponential hazard and a Weibull hazard whose shape parameter,
calibrated from Known Exploited Vulnerabilities catalog timing, captures the
empirical decay of exploitation risk as a vulnerability ages.
Because hazards are additive under independence, per-vulnerability rates
aggregate by summation up to host, network, business unit, and organization.
Candidate remediation actions are simulated and ranked by projected hazard
reduction, giving defenders a defensible, quantitative basis for
prioritization under fixed capacity.
Future work includes extensions for incident likelihood
and financial loss modeling.
\end{abstract}


% ===========================================================================
\section{Introduction}
\label{sec:introduction}

Third-party vulnerability arrival rates outpace organizational remediation
capacity. Over 48,000 Common Vulnerabilities and Exposures (CVEs) were
published in 2025~\cite{cve2025}, a 164\% increase over five
years~\cite{gamblin2025}.
The Forum of Incident Response and Security Teams (FIRST) projects a median of
59,427 CVEs published in 2026, with plausible upper-range scenarios reaching
70,000--100,000 and a 90\% upper bound near
118,000~\cite{first_vuln_forecast_2026,gamblin_cveforecast}.
FIRST's mid-year forecast update reports cumulative CVE publications running
46.3\% above the projected total through April of the original forecast, and
on that basis revises the full-year 2026 projected median to 65,632
CVEs~\cite{first_vuln_forecast_update_2026}.
AI-assisted vulnerability discovery and the structured expansion of the CVE
Numbering Authority (CNA) program are the primary drivers behind this
trend~\cite{first_vuln_forecast_update_2026}.

Historically, organizations remediate between 6.6\% and 27.1\% of their
vulnerabilities per month~\cite{cyentia_p2p}, using various scoring and
prioritization frameworks to guide their efforts, but these frameworks have
significant limitations.
They often rely on static scores that do not reflect the continuous evolution
of threats or the organization's unique context~\cite{allodi_massacci}, and
they lack a direct connection to financial risk, making it difficult for
organizations to understand the potential impact of vulnerabilities and
prioritize accordingly~\cite{woods_bohme}.
More recent frameworks incorporate exploitability and impact factors but stop
short of quantifying the exploitation risk each vulnerability contributes; we
discuss these and other prior approaches in Section~\ref{sec:related}.

The model proposed in this paper is designed to address the vulnerability
arrival rate and remediation capacity incongruity by using exploit likelihood
models (ELMs) such as the Exploit Prediction Scoring System (EPSS) as inputs
into a local Bayesian framework that answers the question:

\begin{quote}
\emph{Given my set of exposed vulnerabilities, how many exploit events should
I expect in a given timeframe, and how can I efficiently reduce that number?}
\end{quote}

The output is a localized exploit hazard model that quantifies exploitation
event arrival rates from individual vulnerability instances to host, network,
business unit, and ultimately organization-wide exploit hazard.
Abstracting exploit risk away from individual vulnerability instances allows
organizations to target exploit risk more holistically through targeted
mitigation actions that reflect the hazard reduced if taken.

\paragraph{Contributions.}
The principal contributions are:
(1)~a hazard-rate formulation that converts ELM outputs into daily exploit
hazard via exponential or Weibull survival models, with per-vulnerability
age-dependent aggregation;
(2)~a simulation function that ranks candidate remediation actions by
projected hazard reduction; and
(3)~a Bayesian framework that allows for control effectiveness updates over
time as controls, or perceptions of controls, change.

% ===========================================================================
\section{Related Work}
\label{sec:related}

In survival analysis, the hazard rate denotes expected events per unit of
time, usually written $h(t)$, where $t$ is time~\cite{klein_moeschberger}.
The same construct appears in reliability engineering as the failure rate of a
component~\cite{rausand_hoyland}, in actuarial science as the force of
mortality~\cite{bowers_actuarial}, and in epidemiology as an incidence
rate~\cite{tenny_boktor}.
Cybersecurity is no different~\cite{soa_cyber}.

A recurring objection to quantitative risk modeling in security is that the
relevant quantities cannot be measured.
Hubbard and Seiersen argue that this objection conflates measurement with
certainty: a measurement is any observation that reduces uncertainty, and a
calibrated estimate expressed as a probability distribution is a measurement
even when no precise count is available~\cite{hubbard_seiersen}.
Linear opinion pooling is a method for aggregating probability distributions
from multiple experts into a single distribution~\cite{clemen_winkler}.

The Common Vulnerability Scoring System (CVSS) classifies vulnerabilities by their qualitative
characteristics and produces an ordinal severity ranking~\cite{cvss}.
To produce the ranking, the qualitative characteristics are assigned numeric
values and an algorithm performs arithmetic with them to produce a score of
how severe vulnerabilities are.
Such arithmetic on ordinal inputs is not valid~\cite{hubbard_seiersen}, and base
scores have been shown to be poor predictors of observed
exploitation~\cite{allodi_massacci}.
CVSS can integrate additional qualitative labels describing threat and
environmental factors, but even with these supplements it remains a
qualitative measure performing arithmetic on labels and ordinal rankings.

Stakeholder-Specific Vulnerability Categorization (SSVC) uses
decision trees that combine qualitative exploitability and impact values to rank
vulnerabilities by which ones require action~\cite{ssvc}.
SSVC has been codified as the primary prioritization method among United States federal
civilian executive branch (FCEB) agencies through adoption by the Cybersecurity and
Infrastructure Security Agency (CISA) via Binding Operational Directive
(BOD)~26-04~\cite{cisa_bod_26_04}.
Outside of FCEB agencies, SSVC adopters must implement their own decision trees
based on their own qualitative judgment to produce a ranking of vulnerabilities that require action.

Exploitation signals fall into two categories: predictive signals that
estimate the likelihood of future exploitation, and observed signals that
record confirmed exploitation events.
Both types are valuable for assessing risk, but they serve different purposes
and have different implications.
The Exploit Prediction Scoring System (EPSS) is an exploit likelihood model (ELM) 
trained on observed exploitation data and publicly publishes a daily probability of 
exploitation for each published CVE in the next 30~days~\cite{epss_dtrap,epss_enhancing}.
Known Exploited Vulnerabilities (KEV)
catalogs~\cite{cisa_kev,vulncheck_kev,circl_kev,euvd_kev} provide disclosure
of observed exploitation for specific vulnerabilities, but often do not reveal
how or where exploitation was detected. As part of BOD~26-04, CISA requires 
federal agencies to remediate KEV-listed vulnerabilities within a specified 
timeframe~\cite{cisa_bod_26_04}.

% ===========================================================================
\section{Exploit Hazard Model}
\label{sec:model}

To get started, we initialize our data set with vulnerability findings to host
mapping, EPSS scores and CVSS attack vector values for each vulnerability, and
a mapping of which controls apply to which hosts on each attack vector.

% ---------------------------------------------------------------------------
\subsection{Measuring Control Effectiveness}
\label{sec:control_effectiveness}

Since EPSS is a global model, we need to quantify how effective our controls
are in preventing exploitation in order to ground it in our local environment.
To do this, a subject matter expert (SME) survey is used with a linear opinion
pool to initialize our collective control exploit prevention effectiveness
prior rate as a probability distribution.
All controls begin with an uninformed $\text{Beta}(1,1)$ prior representing
maximum uncertainty, meaning the controls are equally likely to be completely
effective and completely ineffective, and every value in between.
The uninformed prior does not need to default to $\text{Beta}(1,1)$.
Organizations can choose a different prior distribution based on information
they may already have, but the default minimizes initial bias while maximizing
uncertainty.

\subsubsection{SME Opinion Pool}
\label{sec:sme_pool}

A Linear Opinion Pool aggregates subject-matter expert estimates into an
informative prior.
A survey is conducted to record both their effectiveness estimates and their
self-reported expertise level using a Likert scale~\cite{likert1932}.
The survey elicits two values per SME, a median estimate and a 90th percentile
upper bound, to capture both central tendency and the width of each expert's
uncertainty, allowing the fitted Beta distribution to be asymmetric and take
on the overall shape of the SME's uncertainty.

The expert-weighted responses are modified using the Likert value as a
multiplier to account for their self-reported qualitative expertise ranking, 
and then each pool is used as an input into Powell's hybrid 
method~\cite{powell1970} for finding the root Beta distribution that reflects
each pool. We acknowledge that the arithmetic performed on ordinal Likert values 
here is not valid, but it is used as a practical compromise for weighting 
SME responses to introduce variance in our prior effectiveness estimate.

\begin{figure}[H]
\centering
\includegraphics[width=\textwidth]{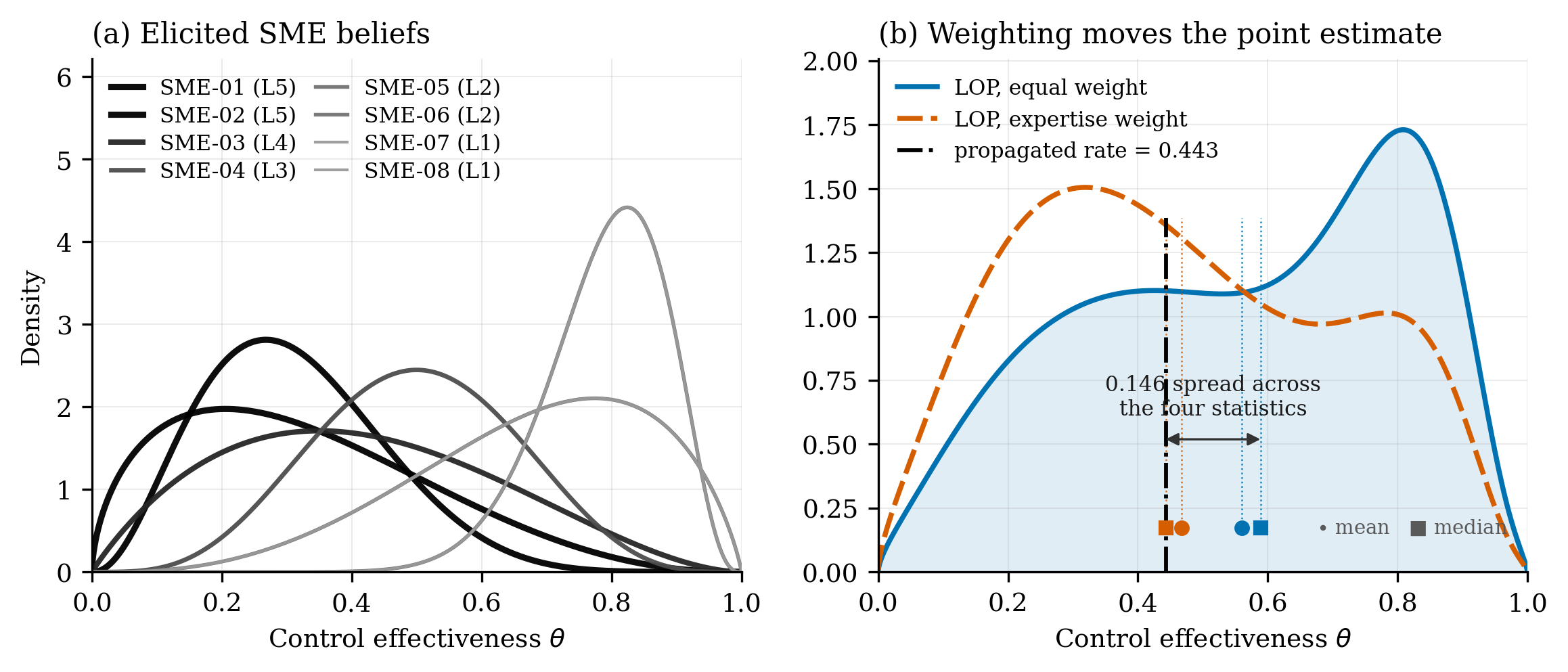}
\caption{\textbf{Pooling expert control-effectiveness estimates.} Linear
opinion pool aggregation of eight subject-matter-expert estimates.
(a)~Per-respondent Beta beliefs recovered from each expert's elicited median
and 90th-percentile pair; line weight encodes self-rated Likert expertise.
(b)~The equal-weight and expertise-weighted linear pools over the same
responses, with the four candidate central tendencies (equal- and
expertise-weighted mean and median) and the conservative minimum the model
propagates.
The pool is a mixture rather than a re-parameterized Beta, so expert
disagreement is preserved rather than averaged away.
Illustrative inputs; the propagated rate depends on the survey responses and
the expertise weighting.}
\label{fig:sme_pool}
\end{figure}

The model uses the lowest central tendency value across each aggregation
method as a conservative point estimate until
observational data narrows the posterior.
In this initial model version, the single point estimate is used as the
control effectiveness rate.
We reserve sampling values from the distribution to
simulate multiple trials as future work in Section~\ref{sec:future}.
Figure~\ref{fig:sme_pool} illustrates the aggregation for a sample
eight-expert survey.

\subsubsection{Local Telemetry Observations}
\label{sec:telemetry}

For controls with observable telemetry, the Beta distribution from the
uninformed prior or SME Opinion Pool updates directly from log
data~\cite{gelman_bda3}:

\begin{equation}
\begin{aligned}
\text{Beta}(\alpha, \beta) &\rightarrow \text{Beta}(\alpha + \text{prevented}, \beta + \text{failed})
\end{aligned}
\label{eq:beta_update}
\end{equation}

However, open-ended log-based observation carries a systematic upward bias.
A control system only counts events it detects. Exploits that evade detection
or do not have signatures deployed are absent from both success and failure
counts.
This is a missing-not-at-random (MNAR) problem~\cite{little_rubin} where the
denominator systematically undercounts failures.
A control that detects 70\% of exploit attempts but is blind to the remaining
30\% will converge toward a posterior near 1.0 rather than its true 0.70
effectiveness.
The SME distribution encodes expert judgment about detection gaps and evasion
techniques that logs cannot capture, serving as a Bayesian anchor that guards
the posterior against detection-biased inflation.
The SME prior introduces conservatism about what the control cannot see, while
observational data introduces precision about what it can see.

\begin{figure}[H]
\centering
\includegraphics[width=0.85\textwidth]{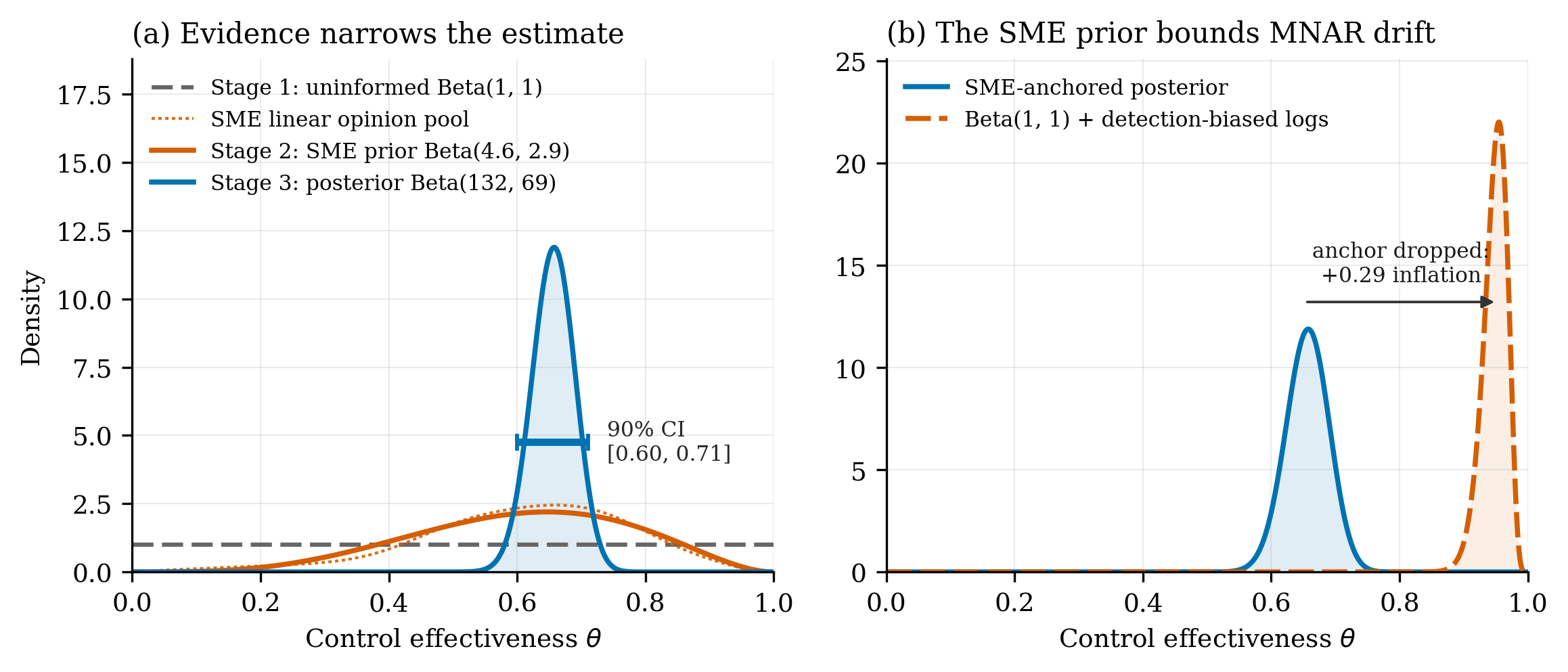}
\caption{\textbf{Bayesian belief convergence.} How the shape of a
control-effectiveness estimate sharpens as evidence accumulates.
(a)~Stage~1 is the uninformed $\text{Beta}(1,1)$ prior representing maximum
uncertainty; Stage~2 incorporates the SME opinion pool, narrowing the
distribution around the experts' central estimate; and Stage~3 reflects
log-based observations, producing a tight posterior whose 90\% credible
interval spans roughly 0.60 to 0.71.
(b)~The SME prior anchors the posterior below the $\approx 1.0$ value that
unadjusted logs would imply for a control blind to a fraction of attempts
(Section~\ref{sec:telemetry}).
Illustrative scenario; actual convergence depends on observation frequency and
event rates.}
\label{fig:convergence}
\end{figure}

Penetration testing, red teaming, and Breach and Attack Simulation (BAS) tools
target the MNAR problem by producing independent, time-bound observations of
control performance.
BAS tools execute simulated attack techniques against control stacks and
record which actions were prevented, detected, or missed, reducing systematic
undercounting of failures.
Red team engagements and penetration tests produce similar observations but
with intelligent technique selection extending beyond BAS libraries.
In both cases, a test executing $n$ attempts and observing $k$ preventions
updates the prior. Under our prior $\text{Beta}(\alpha_0, \beta_0)$, the
posterior is $\text{Beta}(\alpha_0 + k,\; \beta_0 + n - k)$.
The key difference is operational cadence: BAS campaigns run continuously with
broad but scripted technique sets and capture control drift from configuration
changes, while penetration tests run periodically with narrower scope but can
test novel attack chains.
BAS estimates carry their own coverage gap, since simulation libraries test
known techniques, not zero-day or novel evasion, so results represent an upper
bound for known technique classes.
Pairing BAS-calibrated priors with SME regularization for uncovered technique
classes produces posteriors that are empirically grounded where testing exists
and conservatively bounded where it does not.
Both sources feed the same Beta-Binomial update mechanism.
Figure~\ref{fig:convergence} shows how the shape of control effectiveness
uncertainty evolves as observations accumulate between SMEs and local
telemetry.

% ---------------------------------------------------------------------------
\subsection{Applying Control Effectiveness by Attack Vector}
\label{sec:apply_ce}

With a control effectiveness distribution in hand for each control, we combine
it with the per-vulnerability exploit likelihood ($EL_i$) supplied by the ELM
to obtain a control-adjusted exploit likelihood ($EL_{ic}$).
Control effectiveness is applied by attack-vector alignment: a control that
operates at the network layer only affects vulnerabilities whose CVSS attack
vector is Network, a host-based control only affects Local or Physical
vectors, and so on, as read from each vulnerability's CVSS vector string.
This alignment prevents a control from receiving credit against exploitation
paths it does not touch.

For each exploitable vulnerability instance, the probability that exploitation
activity passes through an aligned control is the complement of its prevention
effectiveness, $1 - CE$. Multiplying the ELM likelihood by this bypass
probability yields the residual, control-adjusted likelihood:

\begin{equation}
EL_{ic} = EL_i \times (1 - CE)
\label{eq:elic}
\end{equation}

Applying Equation~\ref{eq:elic} involves two modeling choices. The first is
whether $CE$ enters as a single point estimate or as a draw from the control's
posterior distribution.
We decided on collapsing each control's Beta posterior to its conservative 
minimum-central-tendency estimate of Section~\ref{sec:sme_pool}, and 
propagate a single value.
This keeps the computation cheap and deterministic and is adequate for
ranking, which we verify empirically in Section~\ref{sec:simulation}, but it
discards the posterior's variance.
When multiple controls align to the same attack vector, their effectiveness
rates compose multiplicatively: $EL_{ic} = EL_i \times \prod_{j} (1 - CE_{j})$.
This assumes controls operate independently, where each has its own chance of
preventing an attempt, and one control's outcome does not affect another's.

The second choice concerns how the control is credited against an ELM probability
that may itself summarize many exploitation attempts.
Equation~\ref{eq:elic} treats the ELM-horizon likelihood as a single aggregate
event, so the control gets one effective chance to block it.
If effectiveness were instead applied per attempt, the residual likelihood
would be $1 - (1 - EL_i)^{1 - CE}$, which exceeds $EL_i(1 - CE)$ for
$EL_i, CE \in (0,1)$.
The single-step form can therefore understate residual likelihood
when attempts are relatively independent. We retain it for tractability and
note the per-attempt simulation as an extension in Section~\ref{sec:future}.

% ---------------------------------------------------------------------------
\subsection{KEV Weight}
\label{sec:kev_weight}

The framework includes an optional Known Exploited Vulnerabilities weighting
mechanism.
The weight is a configurable modifier applied to $EL_i$ for any vulnerability
present in a KEV catalog; its default value is 1.0, meaning no effect.
Practitioners may also set a floor that elevates KEV-listed vulnerabilities to
satisfy regulatory requirements such as CISA BOD
26-04~\cite{cisa_bod_26_04}, which carries KEV status forward as a
remediation-timeline criterion, or to reflect operator preference.

We are explicit about what this mechanism does and does not represent, because
it mixes two epistemically different signals.
An ELM produces a predictive probability: the estimated likelihood,
conditional on a vulnerability's features, that exploitation will occur in a
forward-looking window.
A KEV catalog is a realized-observation indicator: a binary record that
exploitation of a specific vulnerability has been confirmed in the wild.
One is a forecast over a population; the other is a recorded outcome for an
individual.
Using KEV membership as a multiplicative modifier on an ELM score therefore
multiplies a probability by an observation, which is not statistically
coherent.
A confirmed-exploited vulnerability has not become ``more probable'' to be
exploited. It has already been exploited, which is a different epistemic
category.
We nonetheless include the mechanism because remediating KEV-listed 
vulnerabilities is frequently a compliance obligation.

% ---------------------------------------------------------------------------
\subsection{Aggregating to the Host}
\label{sec:host_agg}

A host typically carries more than one vulnerability, and we want the
probability that at least one of them is exploited within the horizon.
Assuming exploitation events are independent across the vulnerabilities on a
host, we take the complement of $EL_{ic}$ for each instance (the probability
that it is not exploited), multiply these complements to obtain the
probability that none is exploited, and take the complement of that product to
obtain the host-level grouped exploit likelihood:

\begin{equation}
EL_g = 1 - \prod_{i} (1 - EL_{ic})
\label{eq:elg}
\end{equation}

The independence assumption is the main simplification here.
When several vulnerabilities live in the same software component, exploitation
of one is not independent of the others, and the complement product overstates
$EL_g$; the overstatement grows with the number of correlated vulnerabilities.
For example, fifteen vulnerabilities on the same web framework, each with
$EL_{ic} = 0.02$, yield $EL_g = 1 - (1 - 0.02)^{15} \approx 0.26$ under the
product, whereas treating the component as a single unit by taking the maximum
$EL_{ic}$ yields $EL_g \approx 0.02$, an overestimate of roughly
13$\times$.
The distortion shrinks as vulnerabilities span more independent components.
A component-level grouping is reserved for future work.

% ---------------------------------------------------------------------------
\subsection{From Probability to Hazard}
\label{sec:hazard}

The grouped likelihood $EL_g$ is defined over the ELM's horizon (30~days for
EPSS), but a decision may span a different window, and probabilities do not
scale linearly across horizons (an EPSS value of 0.9 scaled linearly to
365~days would exceed 1.0).
We therefore convert each probability to a daily hazard rate, which does
compose correctly across arbitrary horizons.
The standard probability-to-hazard conversion from survival
analysis~\cite{klein_moeschberger} is:

\begin{equation}
\lambda = \frac{-\ln(1 - P)}{t}
\label{eq:hazard_general}
\end{equation}

Substituting the model's terms, the daily hazard implied by a host's grouped
likelihood over the ELM horizon is:

\begin{equation}
h = \frac{-\ln(1 - EL_g)}{t_{ELM}}
\label{eq:hazard_local}
\end{equation}

\noindent where $t_{ELM} = 30$ for EPSS.
Equation~\ref{eq:hazard_local} produces a single constant hazard per host and
corresponds to the exponential ($k = 1$) case: the instantaneous probability
of exploitation is assumed identical on every day of the horizon, independent
of how long a vulnerability has been present.
Under this assumption the survival function is $S(t) = e^{-ht}$ and the
probability of exploitation over any horizon $T$ recomposes as $1 - e^{-hT}$.

\paragraph{Weibull hazard.}
The constant-hazard assumption implies a vulnerability is as likely to be
exploited on day~1 as on day~300 after disclosure.
Empirically, exploitation activity decays with age as attackers move to newer
targets and defenders patch.
The Weibull distribution~\cite{weibull1951} generalizes the exponential with a
shape parameter $k$ governing how hazard evolves with vulnerability age $t$:

\begin{equation}
h(t) = \frac{k}{\lambda_w}\left(\frac{t}{\lambda_w}\right)^{\!k-1}
\label{eq:weibull_hazard}
\end{equation}

\noindent where $t$ is the number of days since vulnerability disclosure and
$\lambda_w$ is the Weibull scale parameter.
When $k < 1$, hazard decreases with age (exploitation risk decays after
disclosure); when $k = 1$, the model reduces to the constant exponential
hazard; when $k > 1$, hazard increases with age (unlikely for exploitation).
The scale parameter $\lambda_w$ is derived from the ELM probability $p$ and
the baseline window $t_{ELM}$ to preserve consistency with the total
exploitation probability over the ELM horizon:

\begin{equation}
\lambda_w = \frac{t_{ELM}}{\bigl(-\ln(1 - p)\bigr)^{1/k}}
\label{eq:weibull_scale}
\end{equation}

\noindent When $k = 1$, this reduces to $1/h$, the reciprocal of the
exponential rate from Equation~\ref{eq:hazard_local}, confirming that the
exponential model is a special case.

The shape parameter was calibrated from CISA KEV catalog timing data by
computing the interval between NVD publication date and KEV addition date for
each entry, then fitting a Weibull distribution via maximum likelihood
estimation over 1{,}362 usable observations after excluding 131 zero-day
exploits (TTE~$= 0$, a distinct phenomenon) and 131 entries with negative
TTE (exploited before NVD publication).
The fit yields $k = 0.605$; including zero-day exploits (shifted by one
day) yields $k = 0.514$, bounding the shape parameter from both sides.

\begin{figure}[t]
\centering
\includegraphics[width=\textwidth]{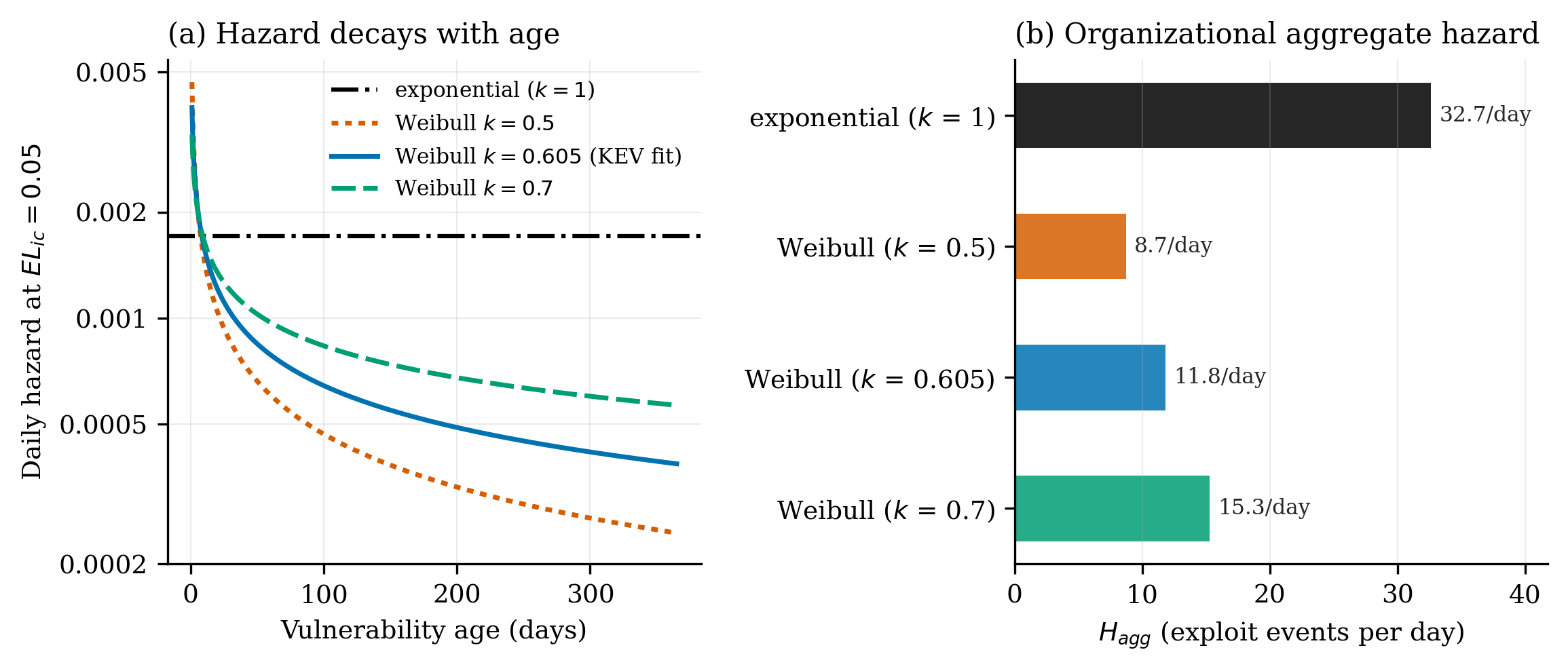}
\caption{\textbf{Exponential vs.\ Weibull hazard.} Both models on identical
inputs. (a)~Daily hazard
against vulnerability age for a fixed $EL_{ic}$; the exponential model is
age-invariant by construction, while the KEV-calibrated Weibull decays.
(b)~Organizational aggregate $H_{agg}$ under each model. The Weibull aggregate falls
below the exponential because the organization's vulnerability population is not
fresh.}
\label{fig:hazard_models}
\end{figure}

Two caveats apply.
First, the CISA KEV addition date is an upper bound on first exploitation, since a
vulnerability may have been exploited before CISA cataloged it.
Second, CISA KEV is a curated catalog reflecting federal priorities rather than a
random sample of exploitation events.
Both biases suggest the true shape parameter is lower (stronger decay) than
the fitted value.
A sensitivity range of $k = 0.5$ to $0.7$ is recommended; at $k = 0.5$, a
vulnerability's hazard at day~90 is roughly one-tenth of its hazard at
day~1, whereas the exponential model treats both days identically.
Figure~\ref{fig:hazard_models} contrasts the two models, both as instantaneous
hazard across vulnerability age and as the organization's aggregate $H_{agg}$.

Under the Weibull model, the hazard rate is computed per vulnerability using
that vulnerability's age, then summed within each asset to obtain the
per-asset hazard rate.
Because hazard rates are additive under independence, this per-vulnerability
summation replaces the complement product in Equation~\ref{eq:elg} when
$k \neq 1$.
Under exponential hazard ($k = 1$), the two are mathematically equivalent;
under Weibull, the per-vulnerability form is required because each
vulnerability's age produces a different instantaneous hazard rate.
The aggregate hazard $H_{agg}$ remains a sum across assets
(Equation~\ref{eq:hagg}) but is no longer constant: it changes as the
vulnerability population ages.
The exploitation probability over a forward-looking window $[t,\, t+T]$ for a
vulnerability of current age $t$ becomes:

\begin{equation}
P(\text{exploit in } [t,\, t\!+\!T])
  = 1 - \exp\!\Biggl(\!-\!\left(\frac{t+T}{\lambda_w}\right)^{\!k}
    + \left(\frac{t}{\lambda_w}\right)^{\!k}\Biggr)
\label{eq:weibull_prob}
\end{equation}

\noindent This conditional probability depends on the vulnerability's current
age, not just the horizon length, reflecting the empirical observation that
older unpatched vulnerabilities carry progressively less exploitation risk
than freshly disclosed ones.

% ---------------------------------------------------------------------------
\subsection{Scaling}
\label{sec:scaling}

The per-host hazard is the model's base unit, and because hazards are additive
under independence it aggregates by summation to any grouping of hosts.
Under the exponential model, where each host's hazard is constant, the
aggregate daily hazard for a set of hosts is

\begin{equation}
H_{agg} = \sum_{i} h_i
\label{eq:hagg}
\end{equation}

\noindent and the same summation over any subset (a network segment, an
application, a business unit, or the organization as a whole) yields that
group's aggregate daily hazard.
Under the Weibull model the aggregate is still a sum over per-vulnerability
hazards, but it is no longer constant. It changes as the vulnerability
population ages, so $H_{agg}$ is evaluated at the horizon's reference age
rather than carried as a fixed rate.

The ability to sum hazards is what lets the model answer its first guiding 
question at any level of the organization. The expected number of exploit 
events over a horizon $T$ is the aggregate hazard integrated over $T$. 
If we treat the aggregate as a Poisson arrival parameter, the probability of at 
least one event is $1 - e^{-H_{agg}T}$.

One consequence is worth drawing out. Fleet size can dominate aggregate hazard. 
A large population of individually low-hazard hosts can contribute more total 
hazard than a small population of high-hazard hosts, a pattern that per-host 
ranking alone would miss and that only becomes visible once hazards are summed. 
Correlated exposure, such as hundreds of hosts behind a single vulnerable
gateway, leaves the expected count unchanged but concentrates events into 
clusters, fattening the tail relative to the Poisson approximation. 
Modeling that correlation through a graph-based dependency structure is reserved 
for future work (Section~\ref{sec:future}).

% ---------------------------------------------------------------------------
\subsection{Simulating Remediation Actions}
\label{sec:simulation}

The hazard aggregates are useful for reporting, but the operational value lies
in identifying which actions reduce the most expected hazard.
We define an action as any discrete change to the model's inputs: remediating
a vulnerability instance, deploying or tuning a control, decommissioning a
host, or updating a control-effectiveness estimate from new telemetry.
Each action modifies one or more inputs and produces a new aggregate hazard.

Actions are grouped by remediation path so that the unit of simulation matches
the unit of work.
A single patch that closes three vulnerabilities on one host is one action; a
firewall rule change that improves network-vector effectiveness across every
host behind it is one action.
For each candidate action $a$, we re-run the pipeline from the modified inputs
forward, holding all other inputs fixed, and record the change in the aggregate 
hazard $H_{agg}$.
Writing $H_0$ for the current metric and $H_a$ for the metric after action
$a$, the risk reduction attributable to $a$ is

\[
  \Delta H_{agg} = H_0 - H_a.
\]

Ranking candidate actions by descending $\Delta H_{agg}$ produces a remediation
queue ordered by expected hazard reduction.
When an effort or cost estimate $c_a$ is available, the queue is instead
ordered by the efficiency ratio $\Delta H_{agg} / c_a$, which maximizes hazard
reduced per unit of remediation capacity.
Because $\Delta H_{agg}$ is defined over the model's inputs rather than any
particular data source, the ranking logic applies to any deployment that can
express its remediation options as input changes, resulting in a feedback loop
that is self-contained and self-consistent.

This ranking is robust to the point-estimate simplification made when applying
control effectiveness (Section~\ref{sec:apply_ce}).
A Monte Carlo experiment ($N$ = 1,000 draws from the $CE$ posteriors of five controls)
confirms that the top-5 remediation ranking is invariant to posterior uncertainty
for moderately informative posteriors (effective sample sizes $\geq$ 20), with
Kendall's $\tau$ = 0.94 between point-estimate and posterior-draw rankings.
The results are consistent under either hazard model 
(exp. $\tau$ = 0.941, Weibull $\tau$ = 0.937).
Instability appears only in the mid-rank region where $\Delta H_{agg}$
values are close.
Propagating full posterior uncertainty through the hazard output on each trial remains a
natural extension for organizations in early deployment with wide posteriors.

\begin{figure}[H]
\centering
\includegraphics[width=\textwidth]{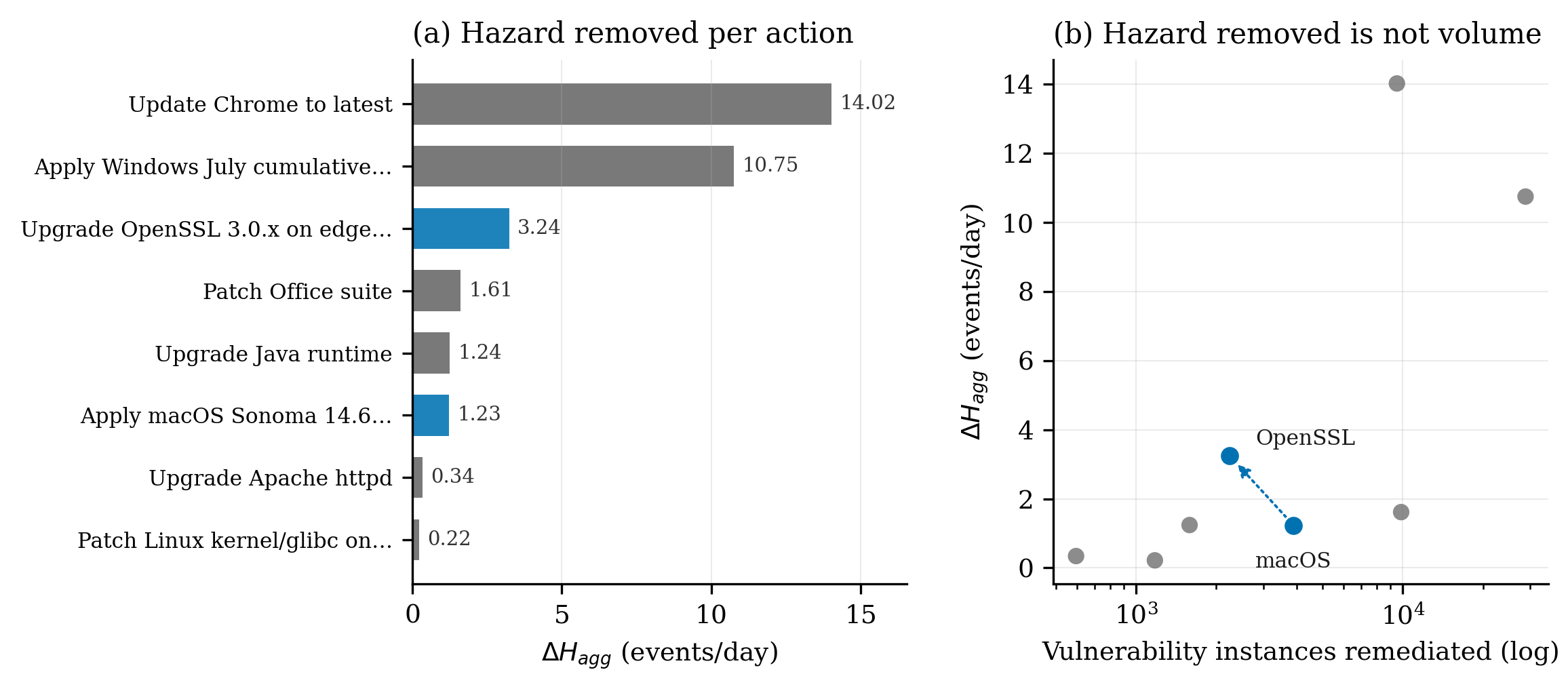}
\caption{\textbf{Hazard reduction is not volume.} Remediation actions ranked
by aggregate daily hazard removed. (a)~$\Delta H_{agg}$ per action. (b)~$\Delta H_{agg}$ against the number of
vulnerability instances each action clears: the OpenSSL upgrade touches fewer
instances on fewer hosts than the macOS update yet removes more hazard, the
inversion a volume- or severity-ranked queue would miss.}
\label{fig:ledger}
\end{figure}

Table~\ref{tab:ledger} shows a remediation ledger for a single simulation
cycle with each candidate action, the aggregate daily hazard it removes
($\Delta H_{agg}$), and the number of vulnerability instances and hosts it
targets.
Figure~\ref{fig:ledger} illustrates why hazard reduction is not a proxy for volume.
In this example, the OpenSSL upgrade clears fewer vulnerability instances 
across fewer hosts than the macOS update, yet removes more aggregate hazard, 
because the library is network-reachable on internet-facing services where 
control effectiveness is lowest and the underlying ELM scores are highest.

\begin{table}[H]
\centering
\caption{\textbf{Remediation ledger.} Illustrative output for one simulation
cycle. Each row is
a single action grouped by remediation path; $\Delta H_{agg}$ is the aggregate
daily exploit hazard removed, ranked descending. Synthetic values on a
$\sim$3{,}000-host organization with a mix of endpoint, server, and edge hosts.}
\label{tab:ledger}
\begin{tabular}{lrrr}
\toprule
Action & $\Delta H_{agg}$ & Vuln.\ instances & Hosts targeted \\
\midrule
Update Chrome to latest (endpoint fleet)   & 14.02 &  9{,}515 & 1{,}240 \\
Apply Windows July cumulative update        & 10.75 & 28{,}887 & 2{,}180 \\
Upgrade OpenSSL 3.0.x on edge services      &  3.24 &  2{,}249 &   410 \\
Apply macOS Sonoma 14.6 security update     &  1.23 &  3{,}901 &   560 \\
Patch Linux kernel/glibc on server pool     &  0.22 &  1{,}177 &   320 \\
\midrule
Total (queued cycle)                        & 29.46 & 45{,}729 & --- \\
\bottomrule
\end{tabular}
\end{table}

% ===========================================================================
\section{Conclusion}
\label{sec:conclusion}

This paper presented a local exploit hazard model: a Bayesian framework that
turns the global exploitation probabilities of an ELM into a defender's own
daily exploit hazard, aggregated from individual vulnerability instances up to
the whole organization.
The model measures control effectiveness as a distribution rather than a
point, applies it to ELM scores by attack-vector alignment, and converts the
result to a hazard rate using standard survival-analysis techniques.
It supports a constant exponential hazard and a KEV-calibrated Weibull hazard
that captures the decay of exploitation risk with vulnerability age, and it
ranks candidate remediations by the hazard they remove.
Its parameters begin as expert-seeded priors and move toward the
organization's own experience as observations accumulate, so the same
machinery serves an organization on day one and after years of telemetry.
The extensions outlined next carry the same hazard output forward into
incident likelihood and financial loss, but the hazard model stands on its own
as a quantitative basis for deciding what to fix first.

% ===========================================================================
\section{Future Work}
\label{sec:future}

\subsection{Extension for an Incident Likelihood Model}
\label{sec:future_incident}

The hazard output answers how often exploitation should occur, not how often
exploitation escalates into an incident the organization would record.
A natural extension maps aggregate exploit hazard to an incident probability
through the Factor Analysis for Information Risk (FAIR) taxonomy's frequency decomposition~\cite{fair_v3}: from
exploit hazard to threat event frequency, and from threat event frequency to
loss event frequency, with a materiality threshold (for public companies, the
``material'' cybersecurity incident of the Security and Exchange Commission's (SEC) Form 8-K Item
1.05~\cite{sec_8k}) separating threat events from the subset that count as
losses.
Treating loss events as a Poisson process gives the probability of at least
one material incident over the horizon as $1 - e^{-\lambda_{LEF}}$, a figure
an organization tracking incidents can validate directly against observed
counts.
Each conversion is a Beta-distributed parameter that starts from an industry
prior and updates from local triage outcomes.
Since EPSS is a global model, the research was limited in bridging the gap 
between global exploitation probabilities and local incident likelihood, but the
FAIR decomposition provides a natural framework for doing so with ELMs trained 
on locally observed exploitation activity.

\subsection{Extension for a Material Risk Model}
\label{sec:future_loss}

Given an incident frequency from the prior section, a further extension 
expresses risk in financial terms.
Drawing per-incident loss from an industry-calibrated log-normal distribution,
parameterized from published median and 95th-percentile losses by sector such
as those in the Cyentia Information Risk Insights Study~\cite{iris2025}, and
combining it with the incident frequency through Monte Carlo simulation
produces a loss exceedance curve.
The log-normal choice captures the right-skew of cyber losses but understates
the extreme tail~\cite{edwards2016}.
As with other parameters, recorded incident costs update the loss
distribution toward the organization's own experience over time.
Further model decomposition of loss magnitude along the FAIR taxonomy
(primary versus secondary loss, and the six forms of loss magnitude) can 
also be incorporated into the Monte Carlo simulation to produce a more 
granular loss exceedance curve.

\subsection{Other Attack Types and a Host Compromise Model}
\label{sec:future_compromise}

The model as presented keys exclusively on vulnerability exploitation, with
controls aligned to CVSS attack vectors.
Much real-world compromise arrives through paths a CVE-centric view does not
enumerate, including non-CVE vulnerability exploitation, phishing and
credential abuse, misconfiguration, and supply-chain exposure.
Any attack vector or attack type could be modeled as its own hazard contribution 
with its own control-effectiveness posterior and folded into the same 
additive aggregate.
A larger extension is a host compromise model that treats initial exploitation
as an entry point rather than the terminal event, using a graph of host and
identity relationships to propagate hazard along lateral-movement paths.
The aggregate would then cover where exploitation is likely to spread as well
as where it is likely to start, and it is the same dependency structure needed
to relax the independent-exploitation assumption behind
Equation~\ref{eq:hagg}.

\subsection{Model Refinements}
\label{sec:future_refinements}

Several narrower refinements follow from simplifications made above.
Introducing a decay function for control effectiveness would capture the reality 
that controls degrade over time, reflecting maintenance cycles, patching delays, 
and evolving threat techniques.
Collapsing each control's Beta posterior to a point estimate before
Equation~\ref{eq:elic} discards posterior variance. A Monte Carlo simulation 
that draws $CE$ from the posterior on each trial, as in
Section~\ref{sec:simulation}, would propagate that uncertainty through to the
hazard output.
Component-level vulnerability grouping, taking the maximum $EL_i$ per
component before the complement product, removes most of the independence
overstatement quantified in Section~\ref{sec:host_agg}.
Applying control effectiveness per attempt rather than per aggregate event,
via simulation over the ELM's implied attempt distribution may address the
understatement of residual likelihood noted in Section~\ref{sec:apply_ce}.
Expanding the KEV catalogs used to calibrate the Weibull hazard would 
reduce the bias from CISA's federal-priority curation.
Finally, using ELMs trained on local exploitation telemetry rather than global 
data would produce grounded $EL_i$ values, and the same model could be applied to
other ELMs and attack vectors with different horizons or feature sets.

% ===========================================================================

\end{document}